\documentclass[12pt,preprint]{aastex}

\slugcomment{To appear in ApJ} \shorttitle{Star Formation
Processes in Outer Disks} \shortauthors{Elmegreen \& Hunter}

\begin{document}

\title{Radial Profiles of Star Formation in the Far Outer Regions of Galaxy Disks}

\author{Bruce G. Elmegreen}
\affil{IBM T. J. Watson Research Center, PO Box 218, Yorktown
Heights, New York 10598 USA} \email{bge@watson.ibm.com}

\and

\author{Deidre A. Hunter}
\affil{Lowell Observatory, 1400 West Mars Hill Road, Flagstaff,
Arizona 86001 USA} \email{dah@lowell.edu}

\begin{abstract}
Star formation in galaxies is triggered by a combination of
processes, including gravitational instabilities, spiral wave
shocks, stellar compression, and turbulence compression. Some of
these persist in the far outer regions where the column density is
far below the threshold for instabilities, making the outer disk
cutoff somewhat gradual. We show that in a galaxy with a single
exponential gas profile the star formation rate can have a double
exponential with a shallow one in the inner part and a steep one
in the outer part. Such double exponentials have been observed
recently in the broad-band intensity profiles of spiral and dwarf
Irregular galaxies. The break radius in our model occurs slightly
outside the threshold for instabilities provided the Mach number
for compressive motions remains of order unity to large radii. The
ratio of the break radius to the inner exponential scale length
increases for higher surface brightness disks because the unstable
part extends further out. This is also in agreement with
observations.  Galaxies with extended outer gas disks that fall
more slowly than a single exponential, such as $1/R$, can have
their star formation rate scale approximately as a single
exponential with radius, even out to 10 disk scale lengths.
H$\alpha$ profiles should drop much faster than the star formation
rate as a result of the rapidly decreasing ambient density.

\end{abstract}


\section{Introduction}

The outer disks of spiral galaxies have a low level of star
formation (Ferguson et al. 1998;  LeLi\`evre \& Roy 2000;
Cuillandre, et al. 2001; de Blok \& Walter 2003; Thilker et al.
2005; Gil de Paz et a. 2005), even though the gas is
gravitationally stable by the Kennicutt (1989) condition.
Triggering by other mechanisms, such as turbulence compression
(Mac Low \& Klessen 2004), supernovae, and extragalactic cloud
impacts (Tenorio-Tagle 1981), might be the reason. As a result,
radial light profiles should not drop suddenly at the stability
threshold, but should taper slowly as various star formation
processes get more and more unlikely and the gas supply
diminishes. The purpose of this paper is to investigate a simple
model of star formation with generalized triggering in a smoothly
varying gas disk. We seek to determine what the overall radial
light profile might be.

The radial light profiles of spiral and irregular galaxies are
typically exponential over 3 to 5 scale lengths (van der Kruit
2001) with rare examples, particularly among low-inclination
spirals, going further (Courteau 1996; Barton \& Thompson 1997;
Weiner et al. 2001; Erwin, Pohlen, \& Beckman 2005; Bland-Hawthorn
et al. 2005). Some galaxies have another, steeper exponential in
the inner disk bulge region (Courteau, de Jong \& Broeils 1996),
which does not concern us here as it may be the result of gas
inflow or bar formation (Kormendy \& Kennicutt 2004). Many
galaxies also have a steep exponential in the far outer disk
(N\"aslund \& J\"ors\"ater 1997; de Grijs, Kregel, \& Wesson 2001;
Pohlen et al. 2002). This outer exponential is the focus of our
discussion. As the outer disk is significantly below the sky
brightness and generally difficult to observe, its properties are
not well known; it may not even be exponential. van der Kruit
(1988) suggested that disk asymmetries can make what is really a
sharp outer truncation appear much smoother when the light
profiles are azimuthally averaged; he noted that very deep
exposures of edge-on disks tend to show sharp edges instead of
smooth outer exponentials.  Florido et al. (2001) showed how a
sharp function could be fitted to outer disk cutoffs. The outer
disk profile also depends critically on the level and uniformity
of the sky brightness subtracted from the image.

The transition from the main disk exponential to the outer disk
profile has several observed characteristics.  The outer disk
scale length is about half that of the inner disk for both spiral
and dwarf irregular galaxies (Hunter \& Elmegreen 2006, hereafter
Paper I). The ratio of the transition, or ``break,'' radius,
$R_{br}$, to the main disk scale length, $R_D$, is 3 to 4 for
spiral galaxies (van der Kruit \& Searle 1981; Barteldrees \&
Dettmar 1994; Pohlen, Dettmar, \& L\"utticke 2000; Schwarzkopf \&
Dettmar 2000; Kregel, van der Kruit \& de Grijs 2002) and $\sim2$
for dwarf and spiral Irregulars (Paper I). There is a slight
increase in this ratio for decreasing $R_D$ among spirals (Pohlen,
Dettmar, \& L\"utticke 2000; Kregel, van der Kruit \& de Grijs
2002; Kregel \& van der Kruit 2004), and another slight increase
for increasing central surface brightness among spirals (Kregel \&
van der Kruit 2004). The first of these two correlations does not
hold for dwarf Irregulars, which have both small disk scale
lengths and small ratios $R_{br}/R_D$. The second correlation does
hold for dwarf Irregulars.  If there is a universal reason for
outer disk transitions (as in the present model), then
correlations which apply to both spirals and irregulars would seem
to be most important. Thus the second correlation, in which
$R_{br}/R_D$ increases with central surface brightness, should be
viewed as fundamental, and the first simply a result of the second
along with the independent correlation between scale length and
central surface brightness found by de Jong (1996) and
Beijersbergen, de Blok, \& van der Hulst (1999).
The apparent ratio $R_{br}/R_D$ should also depend slightly on
galaxy inclination as a result of a tendency to overestimate $R_D$
for edge-on spirals where central extinction flattens the radial profile.

Exponential light profiles in galaxies have been attributed to
several things. Cosmological collapse during galaxy formation,
starting with a nearly uniform spheroid, can produce profiles that
resemble exponentials out to $\sim 2-6$ scale-lengths (Freeman
1970; Fall \& Efstathiou 1980). Exponential disks also arise
through radial flows in viscously evolving disks if the star
formation rate is proportional to the viscosity (e.g., Lin \&
Pringle 1987; Yoshii \& Sommer-Larsen 1989; Zhang \& Wyse 2000;
Ferguson \& Clarke 2001).

Double exponential profiles have no previous explanation (see
review in Pohlen et al. 2004). van der Kruit (1987) proposed that
outer disk truncations arise during galaxy formation and the break
radius is determined by the maximum angular momentum of the
proto-galactic cloud. Kennicutt (1989) suggested that truncation
arises where the gas disk drops below the threshold for
gravitational instabilities.  Elmegreen \& Parravano (1994) and
Schaye (2004) proposed it arises when the ISM converts to a mostly
warm phase, as observed in the outer regions of spirals (Dickey,
Hanson \& Helou 1990; Braun 1997) and dwarfs (Young \& Lo 1996,
1997). Dalcanton et al. (1997), Firmani \& Avila-Reese (2000), Van
den Bosch (2001), Abadi, et al. (2003), Governato et al. (2004)
and Robertson et al. (2004, 2005) simulated galaxy formation with
threshold star formation and obtained exponential profiles with an
outer disk cutoff. None of these models actually obtained double
exponentials, only sharp outer disk truncations.

The theory of disk truncation is highly uncertain, however. The
angular momentum in the outer parts of a galaxy can change over
time during interactions. The gravitational stability threshold
may not be sharp if the ISM cools (Elmegreen 1991) or magnetic
forces remove angular momentum (Kim, Ostriker \& Stone 2002)
during compression. The phase transition may not occur if the
outer gas disk tapers slowly, like $1/R$ (Wolfire et al. 2003).
All of these uncertainties suggest that refined models may
eventually obtain more gradual outer disk truncations.

The presence of double exponentials in dwarf galaxies (Paper I)
places immediate constraints on the models. Most dwarfs have
nearly solid body rotation curves throughout a large fraction of
their optical disks. This means there is little shear, so viscous
evolution should not play a significant role in structuring disk
profiles. There is also no correlation in our Paper I sample
between the break radius and the radius where the rotation curve
changes from near solid body in the inner regions to near flat in
the outer regions. Thus even the outer exponential is not likely
to result from radial migration and evolution related to shear.

Collapse models could in principle be arranged to give the desired
radial profiles, but the collapse models in cosmological
simulations so far have just given inner exponential disks with
relatively sharp outer cutoffs.  There have been no suggestions
yet about how conditions during galaxy formation could be tuned to
give outer double exponentials.  One possibility is that galaxy
collapse gives a single exponential disk and then subsequent
accretion of gas makes the far-outer disk with a different profile
(Bottema 1996). This may explain a sudden decrement in the
rotation speed at the optical disk edge of NGC 4013 (Bottema 1995;
see also van der Kruit 2001), but the decrement could also come
from the prominent warp in that galaxy. If the outer disk is
accreted, then there is no obvious reason why the ratio of outer
to inner scale lengths should be about the same from galaxy to
galaxy, including the dwarfs (Paper I).

Here we consider a star formation model that includes turbulence
and other compressions as cloud formation mechanisms, in addition
to spontaneous gravitational instabilities (see also Kravtsov
2003). Observations of dwarf galaxies have shown that star
formation is not simply regulated by a threshold column density
(see review in Paper I). Star formation clearly occurs in clouds
that stand above the threshold even if the average column density
is below the threshold, and it persists far out in the outer disks
of dwarfs as it does in spirals. It has also become clear that the
ISM in both dwarfs and spiral galaxies is highly structured into
clouds of all sizes, presumably as a result of turbulence and
other processes. For the dwarf galaxies, this conclusion follows
from the log-normal shape of the probability density function of
H$\alpha$ emission (Hunter \& Elmegreen 2004), and from the
power-law power spectra of H$\alpha$ and stellar emissions
(Willett, Elmegreen \& Hunter 2005).  The same power laws for star
formation are seen in spiral galaxies (Elmegreen, Elmegreen, \&
Leitner 2003; Elmegreen et al. 2003). Dwarfs also show power law
or fractal structure in the HI gas, as in the M81 dwarfs
(Westpfahl et al. 1999) and in the Small and Large Magellanic
Clouds (Stanimirovic et al. 1999; Elmegreen, Kim \& Staveley-Smith
2001). The same is observed in local HI (e.g., Dickey et al.
2001). All of these distributions resemble the characteristics of
compressible turbulence as illustrated in simulations (see review
in Elmegreen \& Scalo 2004).

These considerations lead to a model for star formation in a
turbulent, self-gravitating medium.  This model is more general
than the instability model alone as it allows for more processes,
including pressurized triggering of star formation, turbulence
triggering, spiral density wave triggering, and swing-amplified
instabilities.
It should be useful for predictions of outer
disk star formation rates and for semi-analytical models
of star formation in cosmological studies.

\section{Multi-component Model of Star Formation}

Many of the general properties of galaxy disks and star formation
can be combined into a relatively simple model that gives the star
formation rate as a function of radius. These properties lead to
the basic assumptions of the model, as listed here:

\begin{itemize}

\item Galaxies form with a smoothly distributed gas disk having an
outer cutoff, as usually seen in cosmology simulations. This
cutoff will enter the present discussion as the outermost point of
the disk, significantly beyond any break radius that may appear.
We assume in some models that the smooth gas disk is a single
exponential, although other forms will have the same basic
properties.  It will be significant that the star formation rate
takes an approximately double exponential profile even in a gas
disk that is a single exponential. Other models assume an
exponential gas disk with an outer $1/R$ extension. In these
cases, the star formation profile can be a continuous exponential
or a double exponential with the outer part flatter then the inner
part.  In all cases, the star formation profile will drop faster
than the gas profile, but it will rarely truncate suddenly.

\item The ISM is turbulent and partly stirred by pressures related
to existing stars. This means the velocity dispersion tends to
decrease slightly with radius as the stellar disk decreases
exponentially. Such a velocity dispersion decrease is observed for
some spiral galaxies (Boulanger \& Viallefond 1992). Theoretical
discussions of the radial profile of gaseous velocity dispersion
are in Jog \& Narayan (2005). Equating the energy densities of
turbulence and stellar energy input, this gives approximately
$M^2\propto e^{-R/R_D}$ for Mach number $M$, radius $R$, and
stellar exponential disk scale-length $R_D$. The precise form of
this relation is not important to the model; other cases
considered below use a constant Mach number and get about the same
result. The exponential form assumes the gas density for stirring
by supernovae and other stellar pressures is about constant with
radius, as appropriate for the HI medium in the main disks of
galaxies. Then the Mach number alone responds to the stellar
energy density. This is in rough agreement with observations
showing greater HI velocity dispersions for cool HI clouds near
stellar associations (Braun 2005). This equation also emphasizes
that the important Mach number for our model is the one that
regulates the first step of cloud formation, i.e., the conversion
of ambient gas into dense cloud complexes where stars form.  Such
emphasis places turbulence on an equal footing with large-scale
gravitational instabilities. There may be other processes
governing the radial profile of the Mach number inside individual
dense clouds and the Mach number for the mass-weighted ISM as a
whole.

\item The Mach number reaches a minimum value near unity in the
outer disk as a result of either a transition to a warm-dominant
thermal HI phase or a sustained low level of turbulence (Sellwood
\& Balbus 1999).  In either case, cool clouds are still possible
in the compressed regions, but turbulence compression is weak.
Combined with the previous point, this means
\begin{equation}M^2\approx1+Ae^{-R/R_d}\label{eq:mach}\end{equation}
where $A$ is the square of the effective Mach number in the inner
disk. We assume in some models below that $A=100$; the results do
not depend on this value as long as the main part of the inner
disk is Toomre unstable. The most important point for the model is
that some level of turbulence remains in the outer disk so that
turbulence-induced compression makes clouds even where the average
disk is Toomre-stable.  Thus $A=0$ gives acceptable results too.
For the models shown below, Equation \ref{eq:mach} will be used
with either $A=100$ or $A=0$; a more detailed treatment might have the
coefficient $A$ depend on the SFR per unit gas mass,
or on other processes related to interstellar turbulence.

\item Isothermal turbulence produces clouds with a log-normal distribution of
column density, as observed in simulations by Padoan et al.
(2000), Ostriker et al. (2001), and V\'azquez-Semadeni \& Garc\'ia
(2001). Then the probability of a region having a local column
density $\Sigma_g$ is
\begin{equation}P(\Sigma_g)d\ln\Sigma_g=P_0 \exp\left(-0.5\left[\ln
\Sigma_g/\Sigma_p\right]^2 /\sigma^2\right)d\ln
\Sigma_g.\end{equation} The column density at the peak of this
distribution, $\Sigma_p$, will be determined at each radius to
give the appropriate average column density (see below).  The
dispersion of the log-normal may scale with the Mach number of the
turbulence,
\begin{equation}\sigma=\left(\ln\left[1+0.5M^2\right]\right)^{1/2}
\label{eq:disp}\end{equation} as in simulations by Nordlund \&
Padoan (1999).  The log-normal is consistent with the
pixel-to-pixel distribution of H$\alpha$ intensity in Im galaxies
(Hunter \& Elmegreen 2004).

These last two points (with $A>1$) make the ISM more clumpy in the
inner regions than in the outer regions. For the unstable inner
part of the disk, this clumpiness does not matter much for the
star formation rate because it is relatively easy for $\Sigma_g$
to exceed $\Sigma_c$ and also because the instabilities themselves
drive turbulence and cloudy structure. In the stable outer parts,
however, the turbulence-formed clumps and any outward propagating
spirals from the inner disk are the primary regions where
$\Sigma_g>\Sigma_c$ and star formation occurs only in them. This
makes star formation very patchy in outer disks, and it proceeds
at a low average rate.  The rate is not zero even though the
average gas column density, $<\Sigma_g>$, is significantly less
than $\Sigma_c$ because star formation persists in the tail of the
$P(\Sigma_g)$ function.

The log-normal form for $P\left(\Sigma_g\right)$ is not critical
for the double exponential radial profile. It is used here
primarily for convenience and because of its presumed connection
with turbulence. The important point is that
$P\left(\Sigma_g\right)$ has a tail at high $\Sigma_g$ that gets
wider with increasing Mach number, and that some low level of
turbulence compression remains in the gravitationally stable outer
disk.

\item The critical column density for gravity to overcome Coriolis
and pressure forces is the Toomre value appropriate for gas. The
general concept that galaxy edges result from below-threshold
$\Sigma_g/\Sigma_c$ dates back to Fall \& Efstathiou (1980), Quirk
(1972), Zasov \& Simakov (1988) and Kennicutt (1989). We write
$\Sigma_c$ here in terms of the epicyclic frequency $\kappa$ and
the Mach number $M$ instead of the velocity dispersion,
\begin{equation}\Sigma_c=C M\kappa/\left(\pi G\right).
\label{eq:thres}\end{equation} The constant of proportionality, C
(units of velocity dispersion), absorbs the fixed rms speed and
effective adiabatic index that is in the usual expression because
we replaced the dispersion with the radial-varying Mach number. In
our model, $C$ determines where the break radius might occur in
the original exponential; whether it breaks or not depends also on
the run of Mach number with radius.

The instability condition does not indicate only the onset of
swing-amplified or shear instabilities in thin disks, or the onset
of ring instabilities, as originally devised by Safronov (1960)
and Toomre (1964). It is also the condition for the stability of
giant expanding shells (Elmegreen, Palous, \& Ehlerova 2002) and
most likely relevant to the collapse of turbulence-compressed
regions too (Elmegreen 2002). This is because all of these
processes involve gravity, rotation, and pressure. When
$\Sigma_g>\Sigma_c$, gravity overcomes the Coriolis force during
the contraction of the largest cloud that is initially in
pressure-gravity equilibrium. The origin of the cloud does not
matter. If $\Sigma_g<\Sigma_c$, then Coriolis forces disrupt
collapsing spiral arms, expanding shells, turbulence-compressed
clouds, and ISM structures before much star formation begins in
them. Thus, the Toomre condition should be a general condition for
star formation, independent of the detailed triggering processes,
which may be quite varied (Elmegreen 2002). The situation is the
same if the ambient medium cools during the compression, but then
$\Sigma_c$ should be set equal to $C \gamma_{eff}^{1/2}
M\kappa/\left(\pi G\right)$ for effective adiabatic index
$\gamma_{eff}=c^{-2}dP/d\rho$ (Elmegreen 1991), considering
pressure $P$, density $\rho$, and velocity dispersion $c$. Most
likely this occurs in the outer disks where compression can
convert warm HI into cool diffuse clouds. We do not consider this
additional factor here.

\item The star formation rate is proportional to some power of the
{\it local} gas column density $\Sigma_g$ when the threshold is
exceeded. A 1.4 power was observed by Kennicutt (1998) for a wide
range of conditions. A lower limit to the power is $\sim1$, which
also fits the data in some models (Boissier et al. 2003; Gao \&
Solomon 2004).  Note that these power law observations differ
significantly from what one would get for the Toomre instability
alone, where the maximum growth rate,
$\kappa\left(\Sigma_g^2/\Sigma_c^2-1\right)^{1/2},$ increases from
zero rapidly as $\Sigma_g$ begins to exceed $\Sigma_c$, and then
asymptotically levels off to a dependence on $\Sigma_g^1$. This
makes the star formation rate per unit area, which is $\Sigma_g$
times the growth rate, proportional to $\Sigma_g$ raised to a
power greater than or equal to 2. If $\Sigma_g/\Sigma_c\sim1.5,$
for example, then the star formation rate should scale with
$\Sigma_g^{2.8}$. The Toomre condition alone is not appropriate
for star formation because all of the other dynamical processes
that are involved (such as turbulence driven by young stars and
thermal cooling inside compressed regions) change both $\Sigma_g$
and $\Sigma_c$ locally. The Toomre condition assumes an isothermal
uniform gas. If this isothermal assumption is relaxed, then galaxy
disks can become unstable for a wider range of conditions (e.g.,
Elmegreen 1991).  The origin of the observed power law is not
fully understood, but it is probably related to star formation
processes that operate at the local dynamical rate in a medium
that is structured by turbulence (Elmegreen 2002). It should
follow naturally from a full hydrodynamical model that includes
these effects (Kravtsov 2003; Li, Mac Low \& Klessen 2005).

\end{itemize}

These points incorporate the main processes that are believed to
be involved with galactic-scale star formation: ISM turbulence,
pressurized shell formation and other pressurized triggering,
thermal equilibrium, and general gravitational instabilities.
Turbulence and other compressions make the disk cloudy and this
cloudy structure persists in the outer disk even where
$<\Sigma_g>$ is less than $\Sigma_c$, allowing star formation to
continue at large radii. The decrease in the Mach number means the
cloudiness decreases with radius, so the ISM becomes less
turbulent and more smooth in the outer regions, as observed for HI
(Braun 1997). This combination of cloud-forming turbulence with a
Mach number that converges asymptotically to near-unity, along
with cloud-forming instabilities that become decreasingly
important with radius, produces the transition from an inner
near-exponential to an outer near-exponential in our model. The
outer exponential is where the disk is Toomre-stable and the Mach
number is of order unity. Both of these conditions are satisfied
at about the same place when a clear double exponential appears
(see below).

If there are spiral arms in the outer disk, even if they are
generated in the inner disk and radiate dissipatively to the outer
disk, then this model should not change much because these spirals
provide only one more possible source of cloudy structure and
triggered star formation. As long as the gas becomes gradually
less compressive in the outer regions, the star formation rate
tapers off smoothly until the physical edge of the disk (or
ionized edge of the gas) is reached.  Thus the Mach number in our
model should be interpreted as the ratio of rms bulk motion to
sound speed, regardless of whether the bulk motions occur in
spiral shocks, turbulence, or pressurized shells.

Figure \ref{fig-sfr} shows radial profiles of various quantities
from models based on these principles. The models calculate the
results in radial steps of $dR=0.1$ (arbitrary units) for a disk
exponential scale length of $R_D=2.5$  and an outer disk cutoff of
20. At each radius, $R$, the average gas column density is
determined from the initial exponential,
$<\Sigma_g>=\Sigma_{g0}e^{-R/R_D}$, the Mach number is determined
from Equation \ref{eq:mach}, and the dispersion of the probability
distribution function for local column density is determined from
the Mach number using Equation \ref{eq:disp}. Then the peak column
density in this distribution, $\Sigma_p$, is determined from the
integral over $\Sigma_g P\left(\Sigma_g\right)$ by setting the
average column density that results from this integral equal to
$<\Sigma_g>$. The threshold column density, $\Sigma_c$, is also
determined at this radius, from Equation \ref{eq:thres}.

After this setup for the average quantities, the model makes
clouds and determines star formation rates. The local column
density is determined by randomly sampling from the distribution
function $P\left(\Sigma_g\right)$, and then the star formation
rate is set equal to this local column density raised to a power
of 1 or 1.5 in the two cases shown, provided the local column
density exceeds $\Sigma_c$. If the local column density is less
than $\Sigma_c$, then the star formation rate at this position is
set to zero.  To adequately sample the random assignments of
column densities, we consider a number of azimuthal points at each
radius equal to $R/dR$. That is, the size of a cloud is assumed to
be constant with radius. When $R/dR$ random column densities and
resulting star formation rates are determined at each $R$, we
average together these column densities and star formation rates
to give the plotted quantities.

Figure \ref{fig-sfr} shows results for a rotation curve
appropriate for most galaxies (the rotation curve affects only
$\kappa$). This rotation curve is rising in the inner part and
flat in the outer part: $V = V_0 (r/R_D)/ [ 1+ r/R_D]$. The star
formation rate is shown on the left with dashed lines tracing two
exponential profiles to guide the eye. Other quantities for the
same models are shown on the right: critical column density,
$\Sigma_c$ (magenta), Mach number (black dashed), average gas
surface density, $<\Sigma_g>$ (red), and local gas column density,
$\Sigma_g$ (green).  Five cases are considered. In the top two and
bottom two panels, the star formation rate scales with the local
column density to the 1.5 power, while in the middle panel, the
rate scales with $\Sigma_g$ to the first power. The difference is
that when SFR$\propto\Sigma_g^{1.5}$, the inner exponential in
star formation is steeper than the inner exponential in gas;
otherwise the SFR and the gas have the same profiles. The break
radius does not depend noticeably on whether the SFR scales with
$\Sigma$ or $\Sigma^{1.5}$.

The bottom two panels show the difference between models with high
and low critical column densities. When $\Sigma_g/\Sigma_c$ is
lower (bottom panel), less of the disk is unstable and the break
radius is smaller. This is consistent with our observation that
the relative break radius is smaller in dwarfs than in spirals
(Paper I). It occurs because the average surface density is lower
compared to the critical value in dwarfs than in spirals.

The top two panels in Figure \ref{fig-sfr} show cases where the
Mach number has constant values with radius: 1 (top) and 10
(second from top). When the Mach number is 10 throughout, cloud
formation continues at a high rate in the outer part of the disk
(i.e., $\sigma$ in the dispersion of $P(\Sigma_g)$ stays large),
and there is no significant drop in SFR there. Consequently, there
is no clear double exponential.  When the Mach number is 1
throughout, the average SFR profile is almost exactly the same as
for the exponential Mach number (compare to the second panel up
from the bottom which has the same parameters except for the Mach
number), but the rms scatter is much lower when $M=1$ than when
$M\sim10$ in the inner disk. This illustrates how the Mach number
is unimportant for the average star formation rate in the unstable
inner part of the disk. That region is ``saturated'' with star
formation from spiral shocks and instabilities and unable to
produce more star formation even with more compression.  However
the Mach number in the inner disk is important for the detailed
structure of star formation, i.e., for the variability of it and
for the geometry of cloud structure.

These models illustrate how a combination of increasing disk
stability and moderate Mach number can create an approximate
double exponential in the star formation rate when the overall gas
distribution is more uniform. The break radius occurs slightly
beyond the point where $<\Sigma_g>\sim\Sigma_c$ if the Mach number
is of order unity there.  It can occur further out if the Mach
number is still high.

The detailed profile of the star formation rate in regions where
the average column density exceeds the threshold does not depend
much on the rotation curve or Mach number. This is because once
the threshold is exceeded, the threshold no longer enters into the
star formation rate for the simple power law model.

Two other types of radial profiles are found in spiral and
irregular galaxies: those which continue in an exponential fashion
out to the largest measured radius and those which have a
shallower exponential in the outer part (Erwin, Beckman, \& Pohlen
2005; Paper I).

Galaxies of the first type, with a single exponential extending
out to $\sim10$ scale lengths (Weiner et al. 2001; Bland-Hawthorn,
et al. 2005), are difficult to understand with single-component
star formation models because the outer disk should be far below
the Kennicutt (1989) threshold.  Our multi-component model can
reproduce the observation, but the outer gas disk has to fall more
slowly than an extrapolation of the inner exponential. The bottom
panels in Figure \ref{fig-sfr2} show an example.  The average gas
disk is exponential out to $6R_D$ and then it tapers beyond that
as $1/R$ out to $10R_D$. This is consistent with the shallow outer
HI profile in NGC 300 (Puche et al. 1990), which has its stellar
disk extend continuously to $10R_D$ (Bland-Hawthorn et al. 2005).
The profiles of Mach number (using $A=100$) and rotation speed are
the same as in the previous examples, and the local star formation
rate is $\propto\Sigma_g^{1.5}$. In the left-hand panel, the star
formation rate becomes very patchy in the outer part, but the
average rate (solid blue line) follows an overall exponential
profile.  In the right panel, the ratio of $\Sigma_c$ to
$<\Sigma_g>$, which is the Toomre stability parameter, $Q$, is
$5.7$ in the outer regions, indicating great stability on average.
Still, there is a lot of cloud and star formation from local
compressions.

The top part of Figure \ref{fig-sfr2} shows the case if the $1/R$
part of the gas disk begins at a smaller radius, $5.2R_D$, with
all else being the same. Then the profile in the outer disk can be
shallower than in the inner disk.  This is the second type of
profile mentioned above. This explanation differs from that in
Paper I, where we suggested that dwarf galaxies of this second
type, with relatively flat outer parts, could have their
steepening in the central regions because of enhanced star
formation there. This was because the central regions tended to be
bluer than the outer regions; most BCD galaxies were examples of
this. The origin of the flat-outer exponential profiles in barred
S0 galaxies (Erwin, et al. 2005) cannot be due to intense
inner-disk star formation because their Hubble types are too
early. In these S0 galaxies the isophotal contours tend to become
more round with distance beyond the break radius. Radial profiles
determined from deprojected circular averages at all radii could
then introduce false inflections. Other flat-outer profiles in the
Erwin et al. sample have break radii associated with outer rings
or outer Lindblad resonances; these would not be connected with
star formation changes either. More observations of the various
types of radial profiles and their associated gas and star
formation properties are necessary before the relative importance
of these models can be understood.

Our discussion so far has concerned the radial profile of the star
formation rate but not the radial profile of the resulting stars
that form over a Hubble time. Our comparisons between the
predicted star formation profiles and the observed surface
brightness profiles are therefore premature. The next step in this
analysis should be an integration of the star formation rate over
time, but this requires some knowledge of the gas accretion rate,
both as a function of radius and time.  Two limiting cases may be
discussed at this point.  At the end of the star formation
process, after all of the gas has been converted into stars, the
stellar mass profile should reflect the total accreted gas
profile, altered, if need be, by radial gas motions, stellar
migration, minor mergers and tidal stripping.  There is no reason
to believe that this final stellar profile will resemble the star
formation profile.  The second limiting case is at the beginning
of the star formation process, when the outer disk is still
dominated by gas. Then the stellar component is only a small
perturbation to the outer disk and the star formation profile
should be about the same as the accumulated stellar mass profile,
altered again by any accreted stars, radial migrations, and tidal
effects. Fortunately the outer parts of late-type galaxy disks,
beyond $R_{br}$, are usually in this second limit, i.e.,
gas-dominated. For dwarf galaxies, this was shown by Hunter,
Elmegreen \& Baker (1998).  Thus we believe the star formation
profiles derived here for the outer disk are a suitable
explanation for the stellar surface density profiles.

To check this possibility, we ran two of the SFR models shown in Figure
\ref{fig-sfr} (the top panel and the second one up from the bottom)
over a sequence of timesteps. At each timestep and at each radius, we
deducted 1\% of the instantaneous SFR from the average gas mass, and
then added this mass to the integrated star mass. The initial conditions
were the same as in Figure \ref{fig-sfr2}, with no accumulated stars
at first. Figure \ref{fig-evol} shows the resultant SFR, average gas
surface density, and average stellar surface density at three times:
the beginning of the run (blue curves), after 20 timesteps (green
curves), and after 50 timesteps (red curves). We label these timesteps
as 20\% and 50\% of the gas consumption time, respectively, where this
consumption time is considered to be the inverse of the percentage (1\%)
deducted each step. There is no gas accretion from outside the galaxy.
The panels on the left assume the same exponential Mach number profile as
the second panel up from the bottom in Figure \ref{fig-sfr2} (i.e., A=100
in Eq. \ref{eq:mach}) and the panels on the right have the same constant
Mach number profile as the top panel in Figure \ref{fig-sfr2} (i.e., $A=0$
and $M=1$).  Evidently, the gas gets depleted most quickly in the inner
regions, as expected, and the stars build up an exponential profile over
time. The double exponential is still present in the accumulated stars.
The Mach number profile does not matter much, as discussed previously.
At 50 timesteps, the star mass in the steep exponential part of the
outer disk is still much less than the gas mass, in agreement with the
observations cited in the previous paragraph.

We have not commented yet on the relation between star
formation rate and H$\alpha$ surface brightness. The H$\alpha$
surface brightness should become an inadequate tracer of star
formation in the very low-density regions of outer galaxy disks
because the emission measure drops below detectability. The
emission measure through the diameter of a classical Stromgren
sphere with uniform density $n$ is
$\left(6S_0n^4/\left[\pi\alpha\right]\right)^{1/3}$ for Lyman
continuum photon luminosity $S_0$ and recombination rate $\alpha$.
The scaling with $n^{4/3}$ implies that even if the $H\alpha$ flux
does not escape the galaxy, the intensity of an HII region is at
least $e^{16/3}=207$ times fainter at a position four scale
lengths further out compared to the inner disk. This factor
assumes the density is smaller by only the factor $e^{-4}$ without
a corresponding increase in scale height. An increase in scale
height with decreasing disk self-gravity, as $H=c^2/\left[\pi
G\Sigma_T\right]$ for velocity dispersion $c$ and total disk
column density $\Sigma_T$, makes the midplane density roughly
proportional to $\Sigma^2$, in which case the H$\alpha$ intensity
can drop by a factor $e^{32/3}\sim4\times10^4$ in four scale
lengths of an exponential disk. Thus there should be a significant
drop in the radial profiles of H$\alpha$ even with outer disk star
formation of the type discussed here.  This type of $H\alpha$ drop
is in agreement with observations (e.g., Thilker et al. 2005;
Paper I).

The multi-component star formation proposed here also explains
some of the irregularities with the Kennicutt (1989) prediction
that were found in our previous papers on dwarf Irregulars. For
example, star formation in dwarfs often occurs where the average
column density is less than $\Sigma_c$, in contradiction to the
Kennicutt model, as long as there are cool cloudy regions that
locally have $\Sigma_g>\Sigma_c$. This peculiarity was noted
before in many studies of dwarf galaxies (van der Hulst et al.
1993; Taylor et al. 1994; van Zee et al. 1997; Meurer et al. 1998;
Hunter, Elmegreen, \& van Woerden 2001). In the current star
formation model, these cloudy regions form by turbulence and other
processes (pressurized shells, external cloud impacts, end-of-bar
flows, gaseous spirals, etc.) even in sub-threshold regions if the
Mach number for the associated flows is still relatively high. The
revised model also extends the Kennicutt result by allowing for a
typical decrease in velocity dispersion with radius and by
associating the threshold $\Sigma_c$ with any of a variety of
cloud formation processes, and not just isothermal gravitational
instabilities in initially smooth disks.

\section{Conclusions} \label{sec-disc}

Many processes of star formation combine to give the radial
profiles of galaxies. In the inner main-disk regions where the gas
is usually gravitationally unstable in spite of the large Coriolis
and pressure forces, star formation should saturate to its maximum
possible rate. This is the gravitational collapse rate for the
conversion of low density to high density gas, multiplied by the
fraction of the high density gas that is suitable for star
formation, i.e., the fraction in the form of stellar-mass globules
with masses exceeding the local thermal Jeans mass (e.g.,
Elmegreen 2002; Kravtsov 2003). The actual dynamics involved with
the first step, dense cloud formation, will be varied, involving
swing-amplified spiral instabilities, spiral density wave shocks,
compression or shell formation around existing star formation
sites, and turbulence compression. In galaxies with strong stellar
spirals, the spiral shocks may dominate dense cloud formation,
making most clouds spiral-like, as in M51 (e.g., Block et al.
1997), while in galaxies without such strong spirals, another
mechanism should dominate, making most clouds shell-like (as in
the LMC), globular, or hierarchically fractal. In all of these
cases, the same star formation rate per unit area arises, all from
the saturation condition. Thus they all give the Kennicutt-Schmidt
law or something like it in a regular fashion, regardless of the
detailed processes involved.

In the outer parts of disks, some of these processes shut down
completely. There should be no strong stellar spirals beyond the
outer Lindblad resonance for the main (self-amplified) modal
pattern speed, and there should be few swing-amplified stellar
spirals if the Toomre Q parameter is high. Cold cloud formation
should also be more difficult at low ambient pressure. However, a
low level of star formation may sustain itself at large radii by
driving shells and turbulence and by compressing the existing
clouds. Also, gaseous spiral arms can propagate there from the
inner disk, as they are able to penetrate the outer Lindblad
resonance unlike the stellar spirals. Gaseous arms can also form
by instabilities there if there is significant cooling during
compression (because that lowers $\Sigma_c$ through the
$\gamma_{eff}$ parameter). These processes maintain star formation
at levels much lower than the saturation rate given above, and
therefore lower than the Kennicutt-Schmidt law rate, primarily
because an ever-decreasing fraction of the gas can make the
transition from low density to high density in the first step.

In this paper, we modelled all of these processes in a general way
using the few simple rules just mentioned. The transition from
saturated star formation in the inner disk to unsaturated in the
outer disk was followed, and radial profiles were obtained that
look moderately close to real profiles. In the first case
considered, the profile was exponential in the main disk and it
tapered off beyond that with a form that also resembled an
exponential, but steeper for several scale lengths. The ratio of
the break radius to the inner scale length varied with the surface
density (higher surface densities have higher ratios) because more
unstable inner disks have their inner exponentials extend further
out before the transition occurs. This correlation can explain the
observation among both spirals and dwarfs that $R_{br}/R_D$
increases with main disk surface brightness.  In two other cases
with shallower outer gas profiles, the star formation profile
varied between a nearly pure exponential out to $\sim10$ scale
lengths and a shallow outer exponential, depending on where the
transition between the inner and outer gas profiles occurred
relative to the stability threshold radius.  In all cases, the
H$\alpha$ profile should drop much more suddenly with radius than
the star formation profile as the emission measure of individual
HII regions drops rapidly below the detectability limits.

The main ingredients of our star formation model are: a generally
smooth decline of gas column density in the disk with a cutoff in
the far outer part (usually beyond the observations), a turbulent
Mach number that decreases with radius and then levels off to near
unity, or remains near unity throughout, a distribution function
for local column density with a high column-density tail and a
dispersion that increases with Mach number, a column density
threshold for self-gravity to overcome Coriolis and pressure
forces, and a local star formation rate that increases with the
local cloud density when the threshold is exceeded.  For such a
model, the inner exponential occurs where the average column
density exceeds the threshold, almost regardless of Mach number or
Mach number gradient. The outer profile occurs where the gas
column density is sub-critical and the Mach number is relatively
small but non-zero, e.g., near unity. The small but non-zero Mach
number gives turbulence and other dynamical processes the ability
to form clouds that locally exceed the stability threshold, but
these processes are not likely to do this very often. As a result,
star-forming clouds become very patchy in the outer disk, making
the star formation gradient significantly steeper than the gas
gradient.  Only the peaks of the clouds stand above the stability
threshold. Gas cooling during cloud formation is also an essential
ingredient. Cooling to diffuse cloud temperatures and colder is
assumed to follow any significant compression, as predicted
elsewhere based on studies of interstellar thermal equilibrium.

\acknowledgments

We are grateful to the referee for useful comments.  Funding for
this work was provided by the National Science Foundation through
grants AST-0204922 to DAH and AST-0205097 to BGE.

\clearpage

\begin{figure}
\epsscale{0.7} 
\caption{Models of double
exponential disks assuming a single exponential for the average
column density, a probability distribution function for the local
column density based on the results of turbulence simulations,
Mach number variations with radius, column density thresholds for
disk stability using a rising and then flat rotation curve, and
power law star formation rates when the local column densities are
above the threshold. On the left, the highly fluctuating curves
are the resulting star formation rates and the dashed lines are
approximate double-exponential fits to these rates. The curves on
the right show radial profiles of the local gas surface density
(the highly fluctuating curve), the average gas surface density
(straight line), the Mach number (dashed curve), and the threshold
column density (curving smooth line). If the Mach number is a
large constant (second panel from the top), then there is
essentially no double exponential disk. If the Mach number is a
small constant (top panel), or if it decreases with radius to
about unity in the outer regions, then the double exponential
appears in the star formation rate. 
\label{fig-sfr}}
\end{figure}

\begin{figure}
\caption{Models of star-forming disks that extend
far beyond the stability threshold. The average gas column density
profile is exponential in the inner region and $1/R$ in the outer
region. In the top panels, the transition between these two
profiles occurs at $5.2R_D$; in the bottom panels, it occurs at
$6R_D$.  The rotation curve is the same as in the previous figure:
it rises inside $R_D$ and then is flat from $R_D=1$ to 10. The
solid blue lines in the left panels average over the rates (rather
than the logarithms of the rates) so they skim over the tops of
the fluctuating blue curves on these log plots. The red dashed
lines are a guideline to what a pure exponential profile would
look like in this figure.  Depending on where the gas layer levels
off relative to the stability threshold, the star formation
profile can be pure exponential out to $10R_D$ or more (bottom),
relatively flat in the outer regions (top), or rapidly falling in
the outer regions (not shown). \label{fig-sfr2}}
\end{figure}

\begin{figure}
\caption{The evolution of star formation over
time is shown in simple models that assume 1\% of the
instantaneous SFR is subtracted from the gas and added to the
stars at each time step and for each radius. The parameters for
these models are the same as for the second panel up from the
bottom in Fig. 1 (left side here) and the top panel in Fig. 1
(right side here).  The double-exponential radial profile appears
in both the time-integrated stars and the instantaneous SFR.
\label{fig-evol}}\end{figure}

\end{document}